\documentclass[a4paper,10pt]{article}

\usepackage[utf8]{inputenc}
\usepackage[american]{babel}
\usepackage{mathpazo}
\usepackage{graphicx}
\usepackage[font=small,labelfont=bf]{caption}
\usepackage{amsmath}
\usepackage[round]{natbib}
\usepackage{url}

\DeclareMathOperator{\hilbert}{H}
\newcommand{\dd}{\mathrm{d}}
\newcommand{\ii}{\mathrm{i}}

\title{Instantaneous oscillatory direction and phase\\for multivariate timeseries}
\author{Carsten Allefeld}
\date{\small Institute for Frontier Areas of Psychology and Mental Health, Freiburg (Germany)\\
Technical Report, 2008-7-22}

\begin{document}

\maketitle

\begin{abstract} \noindent
This text describes a generalization of the analytic signal \citep{gabor:theory} approach for the definition of instantaneous amplitude and phase to the case of multivariate signals. It was originally written as an appendix for another paper, where the determination of the locally dominant oscillatory direction (the instantaneous amplitude) described here is used as a preprocessing step for another kind of data analysis. The text is reproduced in a `standalone' form because the procedure might prove useful in other contexts too, especially for the purpose of phase synchronization analysis \citep{rosenblum:phase} between two (or more) multivariate \emph{sets} of time series \citep{pascual-marqui:coherence}.
\end{abstract}

\noindent
The local oscillatory behavior of a real-valued univariate signal $x(t)$ is commonly characterized using the corresponding complex-valued analytic signal $z(t)$. It is obtained by combining $x(t)$ with an imaginary part,
$$
z(t) = x(t) + \ii ~ y(t),
$$
which is defined as the Hilbert transform of $x$,
$$
y(t) = \hilbert x(t) = \frac{1}{\pi} \, \textrm{\scriptsize P.V.} \!\!\! \int_{-\infty}^\infty \frac{x(t')}{t - t'} \dd t',
$$
where P.V. denotes the Cauchy principal value of the integral. Under the condition that $x(t)$ is dominated by a single frequency component, its instantaneous amplitude $A(t)$ and phase $\phi(t)$ can be determined via the analytic signal according to
$$
A(t) = |z(t)|,
\quad 
\phi(t) = \arg z(t),
$$
so that
$$
x(t) = A(t) ~ \cos \phi(t)
$$
or
$$
z(t) = A(t) ~ \exp(\ii ~ \phi(t)).
$$
The terms amplitude $A$ and phase $\phi$ as they are used here can be interpreted such that they specify the parameters of a strictly periodic sinusoidal oscillation which \emph{locally matches} the behavior of the observed signal $x(t)$ at a given instant $t$. In particular, $\phi(t)$ attains the value 0 (or equivalently, an integer multiple of $2 \pi$) whenever the actual value of $x(t)$ coincides with the associated instantaneous amplitude $A(t)$.

These properties of the analytic signal can also be utilized to determine the parameters of the locally matching oscillation for a multivariate signal $\vec x(t) = \left ( x_i(t) \right )$ ($i = 1 \ldots K$). We assume that each component signal $x_i(t)$ is dominated by a single frequency and that the frequencies of different signals are practically identical. Using $\vec y(t)$ to denote the channel-wise Hilbert transform of $\vec x(t)$ and $\vec z(t)$ for its channel-wise completion to the analytic signal, the local extension of the signal's oscillatory behavior for instant $t$ is obtained with
$$
\vec z_t(\theta) = \vec z(t) ~ \exp(\ii ~ \theta),
$$
parametrized by $\theta \in [0, 2 \pi]$. Its real part
$$
\vec x_t(\theta) = \vec x(t) \cos \theta - \vec y(t) \sin \theta
$$
gives the multivariate oscillation that locally matches the behavior of the signal at instant $t$; its trajectory is an \emph{elliptical orbit} with conjugate axes defined by the vectors $\vec x(t)$ and $\vec y(t)$.

\begin{figure}
\centering \includegraphics{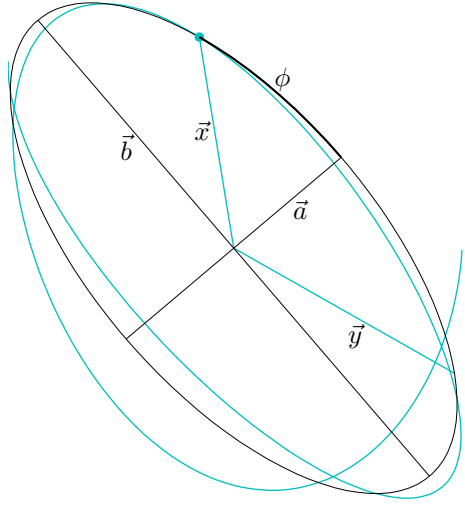}
\caption{Determination of local ellipse axes. The trajectory formed by the multivariate signal is locally matched to an elliptical orbit, which is defined by the data vector $\vec x$ at a given instant and the corresponding vector $\vec y$ from the signal's channel-wise Hilbert transform as conjugate axes. The main axes of the ellipse, $\vec a$ and $\vec b$, are obtained using the associated multivariate instantaneous phase $\phi$.}
\label{ellipse_app}
\end{figure}

From these conjugate axes, the main axes $\vec a(t)$ and $\vec b(t)$ of the local ellipse can be calculated (Fig.~\ref{ellipse_app}). It proves useful to do so via introducing a global (channel-independent) instantaneous phase $\phi(t)$, such that for $\phi(t)$ $\in$ $ \{ 0, \tfrac{1}{2}\pi,$ $\pi, \tfrac{3}{2}\pi \}$ (or equivalents), $\vec x(t)$ coincides with one of the main axis vectors or its negative. This is achieved choosing
$$
\phi(t) = \frac{1}{2} \arctan \frac{2 ~ \vec x(t) \cdot \vec y(t)}{|\vec x(t)|^2 - |\vec y(t)|^2}.
$$
Since the resulting values in the range $[-\frac{\pi}{4}, \frac{\pi}{4}]$ cover only one quarter of a cycle, the outcome may be transformed into an equivalent but more useful representation via a standard ``unwrapping'' procedure (adding or subtracting $\frac{\pi}{2}$ at discontinuity points) to enforce a smooth evolution of $\phi(t)$.

Using this result, the main axis vectors of the locally matching ellipse at instant $t$ are obtained by going backwards along $\vec x_t(\theta)$ by an amount of $\phi(t)$ or forwards by $\frac{\pi}{2} - \phi(t)$, i.e.
$$
\vec a(t) = \vec x_t \left ( -\phi(t) \right )
$$
and
$$
\vec b(t) = \vec x_t \left ( \frac{\pi}{2} - \phi(t) \right ) .
$$
If $\phi(t)$ has been adjusted for a smooth evolution over time, the same can be expected from the resulting $\vec a(t)$ and $\vec b(t)$. It is, however, not clear from this definition which one of these vectors specifies the major and minor axis of the ellipse, respectively, and it is possible that over the course of time the two vectors change roles. For a specific application of this result, further processing may therefore be necessary.

The generalization of the instantaneous phase concept to the multivariate case may be complemented by the definition of a multivariate instantaneous amplitude $\vec A(t)$ such that
$$
\vec z(t) = \vec A(t) ~ \exp (\ii ~ \phi(t)),
$$
which is given by
$$
\vec A(t) = \vec a(t) - \ii ~ \vec b(t).
$$
The channel-wise modulus of this quantity corresponds to the instantaneous amplitudes $A_i(t)$ of the component signals, while the argument comprises the phase differences between the global and the component signal instantaneous phases, $\phi_i(t) - \phi(t)$.

\bibliography{instantaneous}

\end{document}